\documentclass{optica-article}

\journal{opticajournal} 

\articletype{Research Article}

\usepackage{lineno}
\usepackage{amsmath}
\usepackage{hyperref}
\usepackage{physics}

\begin{document}

\title{Estimation of multiple parameters encoded in the modal structure of light}

\author{Alexander Boeschoten,\authormark{1,*} Giacomo Sorelli,\authormark{2} Manuel Gessner,\authormark{3} Claude Fabre,\authormark{1}, Nicolas Treps\authormark{1} }

\address{\authormark{1}Laboratoire Kastler Brossel, Sorbonne Université, ENS-Université PSL, CNRS, Collège de France, 4 Place Jussieu, F-75252 Paris, France\\
\authormark{2}Fraunhofer IOSB, Ettlingen, Fraunhofer Institute of Optronics, System Technologies and Image Exploitation, Gutleuthausstr. 1, 76275
Ettlingen, Germany\\
\authormark{3}Departamento de F\'isica Teòrica, IFIC, Universitat de València, CSIC, C/ Dr. Moliner 50, 46100 Burjassot, València, Spain}

\email{\authormark{*}alexander.boeschoten@lkb.upmc.fr} 


\begin{abstract*} 
We investigate the problem of estimating simultaneously multiple parameters encoded in the shape of the modes on which the light is expanded. For this, we generalize the mode-encoded parameter estimation theory as introduced in Ref.\cite{gessner2023estimation} to a multi-parameter scenario. We derive the general expression for the Quantum Fisher information matrix and establish the conditions under which the multi-parameter Quantum Cramér-Rao bound is attainable.
In specific scenarios, we find that each parameter can be associated with a mode -the detection mode- that is proportional to the derivative of either a single non-vacuum mode or the mean-field mode. For a single non-vacuum mode, the correlation between parameters is determined by the real part of the overlap of these detection modes, while in the case of a strong mean-field by the covariance of the quadrature operators of the derivative modes. In both cases, the attainability of the Quantum Cramér-Rao bound is determined by the imaginary part of the overlap of the detection modes. Our findings provide clear criteria for optimal joint estimation of parameters encoded in the modal structure of light, and can be used to benchmark experimental multi-parameter estimations and find optimal measurement strategies by carefully shaping the modes and populating them with non-classical light.

\end{abstract*}

\section{Introduction}

A fundamental task in physics is to determine the parameters that characterize a specific physical system. Often, light is used as a probe, or light itself holds the parameter of interest. Accordingly, one must identify an effective estimator to extract the parameters from the light, taking into account its noise properties. 

The electromagnetic field of light is defined by its quantum state and the modes on which it is defined \cite{fabre2020modes}. The quantum state can be, for example, a thermal state, a coherent state, or an $ N$-photon Fock state, while the modes can be seen as the classical part of the light, determining the light's shape in both space and time. 

The estimation of a parameter is subject to many sources of noise. Almost all of them can in principle be removed by carefully designing the experiment. However, there is a fundamental quantum noise that can never be eliminated. Quantum metrology, or quantum parameter estimation theory, is the framework to study this fundamental noise. Each type of measurement has its fundamental quantum noise, formulated in the Cramér-Rao bound (CRB) \cite{helstrom1969quantum}, which gives a lower limit to the variance of an unbiased estimator. When optimizing over all possible measurements, one obtains the quantum CRB (QCRB), which essentially quantifies the information about the parameter present in the light. The comparison of the precisions of different types of measurements with the QCRB allows for benchmarking experiments and finding the optimal measurement. 

Often, more than one parameter needs to be estimated, which leads to extra difficulties compared to estimating a single parameter.
Firstly, parameters may be correlated, meaning that the sensitivity of a parameter is lower if a correlated parameter is unknown. These correlations and their impact on the estimation precision are captured by the quantum Fisher information matrix (QFIM) and the associated matrix quantum Cramér-Rao bound.
Secondly, the optimal measurements of different parameters may be incompatible, which means that it is not possible to measure all parameters simultaneously up to the QCRB. In this case, one has to consider the Holevo bound \cite{Hol73}, a tighter bound which is always saturable. While this bound is often hard to calculate, one can show that it is maximally twice as large as the QCRB \cite{DemGorGut2020,TsaAlbDat2020}. The saturability of the QCRB depends on the commutation of the so-called logarithmic derivative operators. When they commute, or even when the expectation value of their commutator is zero, the QCRB coincides with the Holevo bound and is therefore saturable \cite{RagJarDem2016}.

Various expressions for the QFIM are available for different parameters and states \cite{rehacek2017multiparameter,napoli2019towards,zhou2019quantum,Yu2019quantum,wang2021quantum}. The general expression for the QFIM is defined in terms of how the quantum state depends on the parameter via the symmetric logarithmic derivative operator \cite{helstrom1969quantum,liu2020quantum}. Particularly, the single parameter QFI for a parameter $\theta$ is found to be given by the Bures distance between the states $\hat{\rho}(\theta)$ and $\hat{\rho}(\theta+\delta\theta)$\cite{braunstein1994statistical}, and the multi-parameter QFIM is found to be a generalization of this distance \cite{liu2020quantum}. However, in many problems, the parameters do not directly affect the quantum state, but are encoded in the mode of the light, i.e. they affect the spatial and temporal shape, the frequency spectrum, the polarization and the phase of the light \cite{gessner2023estimation,Sorelli_2024}.

Using the quantum theory presented in Ref.\cite{gessner2023estimation}, we study the problem of estimating simultaneously multiple parameters which are encoded in the modal structure of light. A mode parameter change is translated into a change of state in the original mode basis, such that we can find the expression for the QFIM using established techniques. Additionally, we derive an expression for the QFIM in terms of the populated modes and apply it to simple scenarios. We look at two examples, namely the estimation of the parameters which characterize a Gaussian beam and the estimation of the parameters which characterize a Gaussian pulse in a dispersive medium.
With this work, we provide a framework to benchmark multiparameter estimations and find optimal measurements in mode-encoded multiparameter problems.   

\section{Mode parameter estimation theory}
In this work, we are interested in parameters which are encoded in the modal character of the light. To see what these parameters are, let us expand the electromagnetic field operator in an orthonormal mode basis $\{f_k\}$ \cite{fabre2020modes}
\begin{equation}\label{eq:Efieldoperator}
    \vec{\hat{E}}^{(+)}(\vec{r},t)=\sum_k \mathcal{F}_k^{(1)}\hat{a}_k \vec{f}_k(\vec{r},t)
\end{equation}
where $\mathcal{F}_k^{(1)}$ is the electric field per photon and ${f}_k(\vec{r},t)$ is a normalized solution of the Maxwell equations in vacuum, satisfying at any time
\begin{equation}
    (f_m|f_{n})\equiv\frac{1}{V}\int_V d^3 \vec{f}_{m}^*(\vec{r},t)\vec{f}_{n}(\vec{r},t)=\delta_{mn}.
\end{equation}
Here, $\hat{a}_k(\hat{a}_k^\dagger)$ is the annihilation(creation) operator of a photon in mode $f_k$, satisfying $
[\hat{a}_m,\hat{a}_{n}^\dagger]=\delta_{mn}$.

Mode parameters are those parameters that change the modes of the electromagnetic field, but leave the quantum state unchanged. Specifically, a change of these parameters, parametrized by $\vec{\theta}=(\theta_1,\dots,\theta_n)^{T}$, leads to a shifted mode-basis, $\{f_k(\vec \theta)\}_k$. We consider changes of these mode parameters around a value $\vec {\theta}=\vec{0}$, such that $\{f(\vec{0})_k\}_k=f_k$ defines the original mode basis.

To clarify what a mode parameter is, we look at the decomposition of the mode-parameter shifted light field $\vec{ \hat{E}}^{(+)}(\vec{\theta})$. For any parameter which changes the light field, the amplitudes $\hat{a}_k(\vec{\theta})$ of the original modes $\{{f}_k\}$ are parameter independent. 
Mode parameters are defined as those parameters for which we can express the light in a transformed mode basis $\{{f}(\vec{\theta})_k\}_k$, in which we recover the original quantum state: 
\begin{equation}
     \vec{\hat{E}}^{(+)}(\vec{r},t;\vec{\theta})=\sum_k \mathcal{F}_k^{(1)}\hat{a}_k(\vec{\theta}) \vec{f}_k(\vec{r},t;\vec{0})=\sum_k \mathcal{F}_k^{(1)}\hat{a}_k \vec{f}_k(\vec{r},t;\vec{\theta}).
\end{equation}
We see that a mode-parameter change is thus a basis transformation, which can therefore be written as a unitary evolution of the density operator $\hat{\rho}$ defining the state of light: 
\begin{equation}
   \frac{\partial}{\partial\theta_\alpha} \hat{\rho}=-i[\hat{H}_\alpha,\hat{\rho}].
\end{equation}
The generator $\hat{H}_\alpha$ of this transformation was found in Ref.\cite{gessner2023estimation} and has the form

\begin{equation}\label{eq:generatormodeparameters}
    \hat{H}_\alpha=i\sum_{j,k} (f_j|f^\alpha_k)\hat{a}_{j}^\dagger \hat{a}_k=\sum_{k}w_k^{\alpha} \hat{d}_k^{\alpha\dagger}\hat{a}_k
\end{equation}
where $\hat{d}_k^{\alpha\dagger}$ creates a photon in the normalized mode $\tilde{f}_k^\alpha=\frac{i}{w_k^{\alpha}}f_k^\alpha$  with $f_k^\alpha=\frac{\partial}{\partial\theta_\alpha}f_k(\vec{\theta})\big|_{\vec{\theta}=0}$ and
$w_k^{\alpha}=\sqrt{(f_k^\alpha|f_k^\alpha)}$, such that $\hat{d}_k^{\alpha\dagger}=\frac{i}{w_k^{\alpha}}\sum_j(f_j|f_k^\alpha)\hat{a}_j^\dagger$.

To find the ultimate limits on how precisely these parameters can be extracted from a measurement of a state $\hat\rho$, we derive the Quantum Fisher information matrix (QFIM). In the case of multiple unknown parameters $\vec{\theta}= (\theta_1,...,\theta_n)^T$, this QFIM is defined as \cite{helstrom1969quantum,liu2020quantum}:
\begin{equation}\label{eq:mostgeneralQFIM}
    (F_Q)_{\alpha\beta}=\frac{1}{2}\Tr(\hat{\rho}\{\hat{L}_\alpha,\hat{L}_\beta\})
\end{equation}
where $\alpha,\beta$ are the parameter indices and $\{\cdot,\cdot\}$ denotes the anticommutator, and
$\hat{L}_\alpha$ is the symmetric logarithmic derivative operator (SLD) defined by
$
 \frac{\partial}{\partial\theta_\alpha} \hat{\rho}=\frac{1}{2}(\hat{\rho}\hat{L}_\alpha+\hat{L}_\alpha \hat{\rho}).
$
The QFIM can be used to write down a matrix Quantum Cramér-Rao bound, which limits the covariance matrix of the estimators $\vec{\tilde\theta}$:
\begin{equation}
      {\rm cov}(\vec{\tilde\theta})\geq\frac{1}{M}F_Q^{-1}.
\end{equation}
$M$ is the number of repetitions of the measurement, and the covariance matrix is defined as
$
       {\rm cov}(\vec{\tilde\theta})=\left\langle\left(\vec{\tilde\theta}-\langle\vec{\tilde\theta}\rangle\right)\left(\vec{\tilde\theta}-\langle\vec{\tilde\theta}\rangle\right)^T\right\rangle.$

The matrix equality means that ${\rm cov}(\vec{\tilde\theta})-\frac{1}{M}F^{-1}$ is positive-semi definite. Since the diagonal elements of positive-semi-definite matrices are non-negative, the variances of the estimators $\tilde\theta_\alpha$ are bounded by the diagonal elements of the inverse of the FIM:
\begin{equation}
    \Delta^2\tilde{\theta}_{\alpha}\geq \frac{1}{M}(F_Q^{-1})_{\alpha\alpha}\geq \frac{1}{M(F_Q)_{\alpha\alpha}}.
\end{equation}

The last inequality is an equality only if $F_Q$ is diagonal. When the off-diagonal elements are non-zero, there are correlations between the parameters, and the QCRB is higher than in the case of single-parameter estimation.

In contrast to single-parameter estimation, this bound is not always attainable since optimal measurements of different parameters might be incompatible. 
If we are interested in estimating a single parameter, the measurement attaining the QCRB is formed from projectors corresponding to the eigenbasis of the SLD corresponding to the parameter \cite{braunstein1994statistical}. Consequently, in the case of multi parameter estimation,  if SLDs corresponding to different parameters do not share a common eigenbasis, i.e. $[\hat{L}_\alpha,\hat{L}_\beta]\neq0$, the optimal measurements of different parameters do not commute, making this commutation relation a sufficient condition for the multiparameter QCRB to be attainable. However, there is a weaker requirement, stating that only the expectation value of the commutator needs to vanish \cite{RagJarDem2016}. In particular,
\begin{equation}
    \text{Tr}(\hat{\rho}[\hat{L}_\alpha,\hat{L}_\beta])=0
\end{equation}
is a necessary and sufficient condition for the QCRB to be an attainable bound.

We will now present the QFIM and the requirement of attaining this bound for mode-encoded parameters. 
Since mode-encoded parameters lead to unitary evolution of the state, the SLDs can be found in terms of the generators $\hat H_\alpha$ and one readily finds an expression of the QFIM for a state $\hat{\rho}=\sum_a p_a \ket{a}\bra{a}$ \cite{liu2020quantum}:
\begin{equation}\label{eq:GeneralQFIM}
    \begin{split}
        (F_Q)_{\alpha\beta}=2\Tr(\hat{\rho}\{\hat{H}_\alpha,\hat{H}_\beta\})-\sum_{a,b}\frac{8p_a p_b}{p_a+p_b}\Re{\langle a | \hat{H}_\alpha | b \rangle \langle b | \hat{H}_\beta| a \rangle }.
    \end{split}
\end{equation}
Note that for a pure state $\ket{\psi}$ this simplifies to $(F_Q)_{\alpha\beta}=4\text{Cov}_{\ket{\psi}}(\hat{H}_\alpha,\hat{H}_\beta)$.

We can now use the expression for the generators of mode parameters given in \autoref{eq:generatormodeparameters} to obtain the QFIM in terms of the modes of light. A useful form can be found by separating the contribution from the modes of light that are populated, i.e. the non-vacuum modes denoted by $I$, and the modes that are in vacuum, as was introduced in Ref. \cite {gessner2023estimation}. One obtains for a general state $\hat{\rho}$:
\begin{equation}
\begin{split}\label{eq:QFIMmodes}
    (F_Q)_{\alpha\beta}=(F_Q^I)_{\alpha\beta}+4\sum_{j,l\in I}\Re{( f^\alpha_j|\Pi_{\rm{vac}}|f^\beta_l)}\langle \hat{a}_j^\dagger\hat{a}_l\rangle_{\hat{\rho}}.
    \end{split}
\end{equation}
This expression is derived in detail in the supplementary material. We defined $
     \hat{H}^I_\alpha=i\sum_{j,k\in I} (f_j|f^\alpha_k)\hat{a}_{j}^\dagger \hat{a}_k$,
and $
    \Pi_{\rm{vac}}=\sum_{k\not\in I}|f_k)( f_k|$,
the mode projector on the vacuum modes. $F_Q^I$ is the QFIM due to the populated modes, given by \autoref{eq:GeneralQFIM} with $\hat{H}_\alpha=\hat{H}_\alpha^I$. 
The second term in \autoref{eq:QFIMmodes} represents the information contained in the initially unpopulated modes. It arises from the overlap between the derivative modes and the vacuum modes. This contribution scales at most with $\langle{\hat{N}}\rangle$, as expected, since it captures the parameter information that leaks into the vacuum modes and is therefore limited by shot noise.

We now show what requirement we have on the modes for the parameters to be compatible, i.e. for their QCRB to be an attainable bound. First, we express the general requirement in terms of SLDs $\text{Tr}(\hat{\rho}[\hat{L}_\alpha,\hat{L}_\beta])=0 $ in terms of the generators $\hat{H}_\alpha$: 
\begin{equation}\label{eq:attainabilitymixed}
\begin{split}
     \Tr(\hat{\rho}[\hat{L}_\alpha,\hat{L}_\beta])&=4\sum_a p_a\bra{a}[\hat{H}_\alpha,\hat{H}_\beta]\ket{a}\\&-16\sum_{a,b}\frac{p_a^2p_b}{(p_a+p_b)^2}\left(\bra{a}\hat{H}_\alpha\ket{b}\bra{b}\hat{H}_\beta\ket{a}-\bra{a}\hat{H}_\beta\ket{b}\bra{b}\hat{H}_\alpha\ket{a}\right).
\end{split}
\end{equation}
For a pure state, the last term vanishes, and we find a requirement in terms of the populated modes:
\begin{equation}\label{eq:attainabilitypurestate}     \langle [\hat{H}_{\alpha},\hat{H}_\beta]\rangle=\sum_{i,j}\left((f_i^\alpha|f_j^\beta)\langle\hat{a}_i^\dagger \hat{a}_j\rangle- (f_j^\beta|f_i^\alpha)\langle\hat{a}_j^\dagger \hat{a}_i\rangle\right)=
     2i \Im{\sum_{i,j\in I}(f_i^\alpha|f_j^\beta)\langle\hat{a}_i^\dagger \hat{a}_j\rangle}.
\end{equation}
First of all, we see that if the derivative modes are orthogonal, the QCRB is attainable, i.e. the optimal measurements are compatible. On the other hand, in cases where $\langle\hat{a}_i^\dagger \hat{a}_j\rangle\in\mathbb{R}$, the imaginary part of the overlap of different derivative modes determines the compatibility, which has an interpretation in terms of the quadratures of detection modes, as we will comment on in the next two sections.

\section{Single mode}
Let us now look at the case where only a single mode is occupied and all remaining modes are in the vacuum, i.e. $I=\{f\}$. We parametrize a general mode as $f(x;\vec{\theta})=A(x;\vec{\theta}) e^{-i\phi(x;\vec{\theta})}$, where $A(x;\vec{\theta})\in\mathbb{R}_{>0}$ determines the amplitude or spatial and temporal shape of the mode, while $e^{-i\phi(x;\vec{\theta})}$ is the phase of the mode. 
Before looking at the QFIM and the attainability of the QCRB, we provide some intuition by looking at the evolution of a general single-mode pure state for a change in the parameters:
\begin{equation}
    |\psi\rangle_{f(\vec{\theta})}=U(\vec{\theta}) |\psi\rangle_{f}
\end{equation}
with $U(\boldsymbol{\theta}) = \prod_{j=1}^{n} e^{-i \theta_\alpha \hat{H}_\alpha}$. For a small change in the parameters, we find
\begin{equation}
     |\psi\rangle_{f(\vec{\theta})} \approx (1 - i \sum_{\alpha} \theta_\alpha  \hat{H}_\alpha)|\psi\rangle_{f}=(1+\sum_\alpha\theta_\alpha w^\alpha\hat{d}^{\alpha\dagger} \hat{a})|\psi\rangle_{f}.
\end{equation}
where $\hat{a}$ annihilates a photon in the only populated mode $f$. We see that for each parameter  $\hat{d}^{\alpha\dagger}$ creates a photon in the mode $\tilde f^\alpha=\frac{i}{w^\alpha} f^\alpha$, where $w^\alpha=\sqrt{(f^\alpha|f^\alpha)}$ gives the sensitivity of each parameter. These are the modes containing the information about the parameters, and we call them the detection modes. 

For single-mode problems, we have $\hat{H}^I_\alpha=i(f|f^\alpha)\hat{N}$ and the general formula for the QFIM for mode parameters \autoref{eq:QFIMmodes} reduces to a simple form
\begin{equation}\label{eq:singlemodeQFIM}
    (F_Q)_{\alpha\beta}=4(f^\alpha |f)(f|f^\beta )F_Q^I+4\Re{(f^\alpha|f^\beta)-(f^\alpha|f)(f|f^\beta )}N
\end{equation} 
where $N=\langle\hat{N}\rangle_{\hat{\rho}}$ is the mean number of photons in mode $f$. $F_Q^I$ is given by \autoref{eq:GeneralQFIM}, replacing  $\hat{H}_\alpha,\hat{H}_\beta$ by $\hat{N}$, explicitly
\begin{equation}\label{eq:QFIMfirstterm}
    F_Q^I=4\Tr(\hat{\rho}\hat{N}^2)-8\sum_{a,b}\frac{p_ap_b}{p_a+p_b}|\langle a|\hat{N}|b\rangle|^2,
\end{equation} which reduces for a pure state to $F_Q^I=4\Delta^2\hat{N}$.
We will now look at two types of mode parameters for which the QFIM takes a particularly simple form in which the role of the derivative mode becomes clear. 

First, if the parameters are only encoded in the amplitude of the mode but not in the phase $\phi$, i.e. $f(x;\vec{\theta})=A(x;\vec{\theta}) e^{-i\phi(x)}$, we have $(f|f^\alpha)= 0$. This is the case, for example, in the estimation of the transverse displacement of a Gaussian beam \cite{Treps:03} or the Mach-Zehnder interferometer \cite{gessner2023estimation}. In this case, $H^I_\alpha=0$ and the first term in the QFIM vanishes, leading to a simple expression for the QFIM:
\begin{equation}
    (F_Q)_{\alpha\beta}=4\Re{(f^\alpha|f^\beta)}N.
\end{equation}
This shows that the real part of the overlap of the derivative modes $\Re{(f^\alpha|f^\beta)}$ determines the amount of correlation between the parameters, given by the off-diagonal elements of the QFIM. It also shows that the precision is always limited by the shot noise $N$, so any improvement below the Standard Quantum limit needs a second populated mode. This has been done, for example, for the transverse displacement of an HG$_{00}$ Gaussian-shaped beam \cite{Treps:03} and the Mach-Zehnder or Michelson-Morley interferometer \cite{LIGO2013}. 

Secondly, if the parameters are encoded only in the phase of the mode and the phase is independent of $x$, i.e. $f(x;\vec{\theta})=A(x) e^{-i\phi(\vec{\theta})}$, then
$\partial_\alpha f=-i(\partial_\alpha \phi({\vec{\theta}})) f$. This means that the derivative modes are up to a phase equal to the initially populated mode. They thus have no overlap with the vacuum modes, and the second term in the QFIM is zero. This is the case, for example, with the estimation of orbital angular momentum \cite{ambrosio2013}. The QFIM in this case reduces to
\begin{equation}
    \begin{split}
(F_Q)_{\alpha\beta}=4(f^\alpha|f^\beta)F_Q^I.
    \end{split}
\end{equation}  
We see again that the overlap of the derivative modes (which is always real in this case) determines the amount of correlation between parameters. In this case, it is also possible to obtain quantum enhancement by, for instance, using squeezed light.

We now look at the attainability of the QCRB for a single populated mode. The second term of \autoref{eq:attainabilitymixed} is in this case always zero, since we have $\bra{a}\hat{H}_\alpha\ket{b}\bra{b}\hat{H}_\beta\ket{a}=(f|f^\alpha)(f|f^\beta)|\langle a|\hat{N}|b\rangle|^2=\bra{a}\hat{H}_\beta\ket{b}\bra{b}\hat{H}_\alpha\ket{a}$. Therefore, we are left with the first term which simplifies for a single mode to $\sum_a p_a\Im{\sum_{jm}(f_j^\alpha|f_m^\beta)\langle a|\hat{a}_j^\dagger\hat{a}_m|a\rangle} = \sum_a p_a\Im{(f^\alpha|f^\beta)\langle a|\hat{N}|a\rangle }$.  Since $p_a\in\mathbb{R}$ and  $\langle a|\hat{N}|a\rangle\in \mathbb{R}_{>0}$ for non-vacuum states, we arrive to the requirement
\begin{equation}\label{eq:attainabilitySingleMode}
    \text{The QCRB is attainable if and only if }\Im{(f^\alpha|f^\beta)}=0.
\end{equation}
We see thus that the compatibility is determined by the imaginary part of the overlap of the derivative modes.
We can relate this to the quadratures of the detection modes by using the relation $[\hat{d}^\alpha,\hat{d}^{\beta\dagger}]=(\tilde f^\alpha|\tilde f^\beta)$ \cite{fabre2020modes}, such that 
\begin{equation} \label{eq:imaginaryoverlapquadratures}
   2i\Im{(f^\alpha|f^\beta)}/w^\alpha w^\beta=[\hat{d}^\alpha,\hat{d}^{\beta\dagger}]-[\hat{d}^\beta,\hat{d}^{\alpha\dagger}]= [\hat{q}^\phi_\alpha,\hat{q}^\phi_\beta]
\end{equation}
with $\hat{q}^\phi=e^{-i\phi}\hat{d}^\alpha+e^{i\phi}\hat{d}^{\alpha\dagger}$ being the generalized quadrature operator of the detection modes $\tilde{f}^\alpha$.
If the imaginary part of the overlap of the derivative modes is non-zero, the quadratures of the detection modes do not commute. This means they are conjugate variables, corresponding to orthogonal quadratures of the same mode.

\section{Intense mean field}
A commonly appearing situation in experiments is when a certain mode $f_0$ is prepared with an intense mean field $\langle\hat{a}_0\rangle=\sqrt{N_0} $ which without loss of generality we choose to be real, while the other modes have low photon numbers, i.e. with $N_0\gg 1$. We call this mode the \textit{mean-field} mode, since $f_0=\frac{1}{\sqrt{N_0}}\langle \hat{E}^{(+)}\rangle$. If we further assume that the parameters are encoded in the amplitude of the mode, such that $(f_0|f_0^\alpha)=0$, as is the case in the Mach-Zehnder interferometer or the measurement of transverse beam displacement, we find for the generator of the mode parameter:
\begin{equation}
    \hat{H}_\alpha= i\sum_{j>0}(f_j|f_0^\alpha)\hat{a}_j^\dagger\hat{a}_0+h.c.+i\sum_{j,k>0}(f_j|f_k^\alpha)\hat{a}_j^\dagger\hat{a}_k,
\end{equation}
where $h.c.$ denotes the hermitian conjugate. We introduce the operator $\delta\hat{a}_0$, which describes fluctuations of the field around its mean value $\alpha_0$: $\hat{a}_0=\langle\hat{a}_0\rangle +\delta\hat{a}_0$. We can then approximate the generator  as
\begin{equation}
    \hat{H}_\alpha=\sqrt{N_0}w^\alpha \hat{q}_\alpha,
\end{equation}
where $\hat{q}_\alpha= \hat{a}_\alpha+\hat{a}_\alpha^\dagger $ are the amplitude quadrature operators of the normalized derivative modes $\tilde{f}_0^\alpha=\frac{i}{w^\alpha}f^\alpha_0$ with normalization constants $w^\alpha=\sqrt{(f_0^\alpha|f^\alpha_0)}$. As we will see, these modes contain the most information about the parameters $\alpha$ and are therefore the detection modes. Since we assume a strong mean field, we can omit the $O(\delta\hat{a}_0)$ terms and the contributions from other modes than the mean-field mode, and then using \autoref{eq:GeneralQFIM}, we find for the QFIM
\begin{equation}
    (F_Q)_{\alpha\beta}= 4N_0w^\alpha w^\beta \left( \frac{1}{2}\Tr\left(\hat{\rho}\{\hat{q}_\alpha,\hat{q}_\beta\}\right)-\sum_{a,b}\frac{2p_a p_b}{p_a+p_b}\Re{\langle a|\hat{q}_\alpha|b\rangle\langle b |\hat{q}_\beta|a\rangle}\right)
\end{equation}
which reduces for a pure state $\hat{\rho}=|\psi\rangle\langle\psi| $ to 
\begin{equation}(F_Q)_{\alpha\beta}=4N_0w^\alpha w^\beta  \text{Cov}_{\ket{\psi}}(\hat{q}_\alpha,\hat{q}_\beta).
\end{equation} 
This shows that a higher precision on the parameters can be achieved for a low variance of the quadratures of the detection modes, which can be obtained by squeezing. On the other hand, if there are correlations between the quadratures of different modes, this leads to a correlation between the parameters, i.e. a reduced sensitivity on parameter $\alpha$ if the correlated parameter $\beta$ is unknown. We see that when the detection modes are non-orthogonal, the quadratures are always correlated, leading to correlated parameters. 

We also look at the compatibility of optimal measurements in this case. From \autoref{eq:attainabilitypurestate} we find that for attaining the QCRB in a pure state we need $
    \langle[\hat{H}_\alpha,\hat{H}_\beta]\rangle=N_0w^\alpha w^\beta  \langle[\hat{q}_\alpha,\hat{q}_\beta]\rangle=0$. Similarly as in the single-mode case \autoref{eq:imaginaryoverlapquadratures}, we can translate this into a requirement on the modes and find
\begin{equation}\label{eq:compatibilityMeanfield}
    w^\alpha w^\beta[\hat{q}_\alpha,\hat{q}_\beta]
   =2i\Im{(f_0^\alpha|f_0^\beta)}.
\end{equation}
Since $N_0>0$, we find, as in the single-mode case, that the QCRB is attainable if and only if the imaginary part of the overlap of the derivative modes vanishes, or equivalently, when the quadratures of the detection modes  commute.
The reason for this becomes more clear if we expand the electric field operator for small values of the mode parameters:
\begin{equation}
    \begin{split}
        \hat{{E}}^{(+)}({\vec{\theta}})&=\sqrt{N_0}f_0(\vec{\theta})+\sum_{i}\delta\hat{a}_if_i\\
        &=\sqrt{N_0}(f_0+\theta_\alpha f_0^\alpha+\theta_\beta f_0^\beta+\cdots )+\sum_{i}\delta\hat{a}_i f_i,
    \end{split}
\end{equation}
where for convenience we omitted the single-photon fields $\mathcal{F}_i^{(1)}$. We see that the information about the parameters is encoded in the amplitude quadratures of the derivative modes, and measuring the field quadrature with homodyne detection is in fact the optimal measurement in the case of single-parameter estimation \cite{delaubertThesis07,Delaubert:06}. If the derivative modes are orthogonal, \autoref{eq:compatibilityMeanfield} is zero and the QCRB is achievable since the quadrature measurements can be done independently as they correspond to different modes. However, if the derivative modes are non-orthogonal, i.e. $(\tilde{f}_0^\alpha|\tilde{f}_0^\beta)=\Delta=\Delta_R+i\Delta_I$ with $\Delta_I,\Delta_R\in\mathbb{R}$, the real ($\Delta_R$) and imaginary part ($\Delta_I$) of the overlap determine the correlation and the compatibility. To see this, we use the Gram-Schmidt method to orthogonalize the modes and obtain
\begin{equation}
    \hat{{E}}^{(+)}({\vec{\theta}})=\sqrt{N_0}\left[f_0+(\theta_\alpha w^\alpha+(\Delta_R+i\Delta_I)\theta_\beta w^\beta)f_1+\sqrt{1-|\Delta|^2}\theta_\beta w^\beta f_2
    \right]+\sum_{i}\delta\hat{a}_i f_i,
\end{equation}
where we choose the mode basis $\{f_k\}$ such that $f_1=-i\tilde{f}_0^\alpha$ and  
$f_2=-i(\tilde{f}_0^\beta-\tilde{f}_0^\alpha\Delta )/\sqrt{(1-|\Delta|^2)} $, which are orthonormal modes.
When we compare this with the expansion of the field in terms of quadrature operators, i.e. $\hat{{E}}^{(+)}=\frac{1}{2}\sum_n(\hat{q}_n+i\hat{p}_n)f_n$, we see that 
\begin{equation}
    \begin{split}
        \theta_\alpha w^\alpha+\Delta_R\theta_\beta w^\beta&=\frac{1}{2\sqrt{N_0}}\langle\hat{q}_1\rangle,\\
       \Delta_I\theta_\beta w^\beta &=\frac{1}{2\sqrt{N_0}}\langle\hat{p}_1\rangle.   
    \end{split}
\end{equation}
This shows that for $\Delta_R\neq0$, measuring the $\hat{q}$-quadrature of mode $f_1$ leads to a combined measurement of $\theta_\alpha$ and $\theta_\beta$, i.e. these parameters are correlated. For $\Delta_I\neq0$, estimating both parameters optimally requires measuring the $\hat{q}$ and $\hat{p}$ quadratures of the same mode $f_1$, which are conjugate variables, and are therefore incompatible. There is still information in the $\hat{q}$ quadrature of $f_2$, but this information diminishes for larger overlap $\Delta$.

\section{Applications}

\subsection{Characterization of a Gaussian transverse mode}

\begin{figure}
    \centering
    \includegraphics[width=\linewidth]{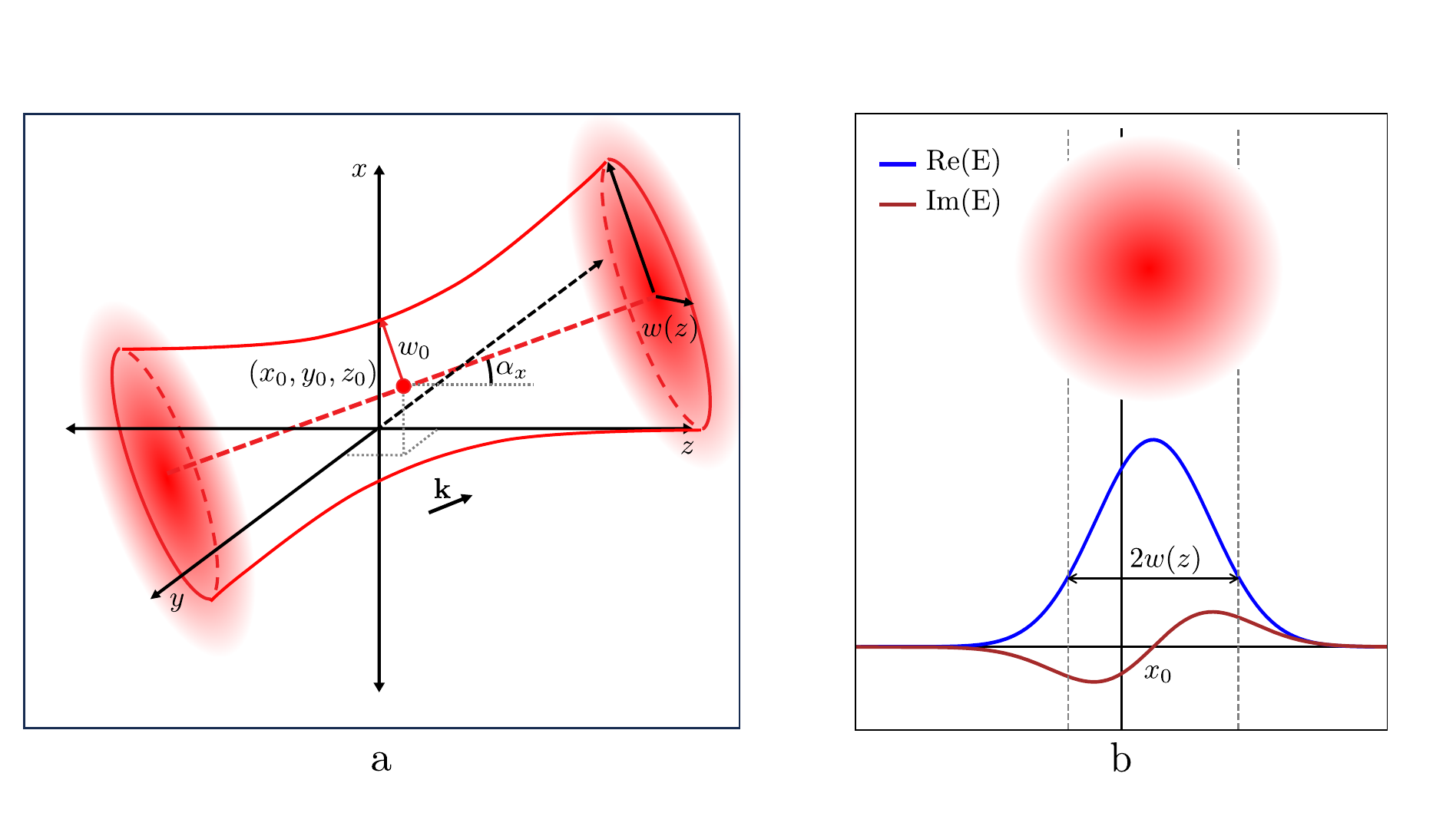}
    \caption{(a) A Gaussian beam with waist $w_0$ at its origin $(x_0,y_0,z_0)$ propagating along the $z$ axis with a small tilt $\alpha_x$ in the $xz$-plane. (b) The real (blue) and the imaginary (brown) part of the tilted and displaced Gaussian transverse mode at a position $z$.}
    \label{fig:gaussianbeam}
\end{figure}

The characterization of a Gaussian transverse mode is an example of the estimation of multiple parameters in a single mode of light.
A Gaussian beam which propagates in the $z$-direction is characterized by the position of its waist $(x_0,y_0,z_0)$, i.e. the position where $w(z)=w_0$, the waist size $w_0$ and the wavelength or wave number $k=\frac{2\pi}{\lambda}$ and a possible tilt in the reference frame, as illustrated in \autoref{fig:gaussianbeam}.
We assume the tilt of the beam $\alpha_x$ or $\alpha_y$ in the $x-z$ and $y-z$ plane, respectively, to be small. 
We are interested in estimating small variations of $(x_0,y_0,z_0,w_0,\alpha_x,\alpha_y)$ around their initial values $(0,0,0,w_0,0,0)$. Furthermore, we choose for convenience $z=0$, i.e. we consider focused transverse Gaussian modes (we can do the analysis for any value of $z$, leading to the same results) and we consider monochromatic light, i.e. we are not interested in estimating $k$. Calculating the six derivative modes and using \autoref{eq:singlemodeQFIM}, one finds for the QFIM
\begin{equation}
    F_Q(\hat{\rho}) =N
\operatorname{diag} \left[
\frac{4}{w_0^2} , 
\frac{4}{w_0^2} , 
\frac{1}{z_R^2} \left(1+\frac{F_Q^I}{N} \right),
\frac{4}{w_0^2}, 
k^2 w_0^2 , 
k^2 w_0^2 
\right].
\end{equation}
For more details, see the supplementary material. Here, $F_Q^I$ is given for a general state by \autoref{eq:QFIMfirstterm} and $N=\langle\hat{N}\rangle_{\hat{\rho}}$. We observe that the QFIM is diagonal, meaning that the parameters are all uncorrelated. The result agrees with what was found for the localization of the waist in \cite{wang2021quantum}. If one has the possibility to measure the initial phase of the light, there is also a contribution from the carrier $e^{-ikz}$, which gives an extra contribution to the derivative mode $f^{\theta_3}$, as discussed in \cite{Boucher:17}. This leads to a small modification of the QFIM, namely that $(F_Q)_{\theta_3\theta_3}=\left((2\alpha^2-1)/z_R^2\right)\left(F_Q^I + N\right)$, with $\alpha=1-(kw_0)^2$.

If we look at the attainability of the QCRB, we note that not all optimal measurements are compatible. First, detection modes associated to tilt and displacement estimation are proportional but in quadrature and thus their imaginary overlap is equal to 1, leading to an unattainable QCRB. This was also found in \cite{xia2023toward}. For waist size and axial displacement estimation, detection modes are neither proportional nor linearly independent, with a non-zero imaginary overlap, and thus here also the QCRB is unattainable. These are clear manifestations of Heisenberg's uncertainty principle, since the detection modes correspond to spatial conjugate variables \cite{Delaubert:06}.

\subsection{Phase velocity, group velocity and group-velocity dispersion in a Gaussian pulse}
A second example, where we see a correlation between different parameters, is the characterization of a pulse of light in a dispersive medium. The parameters $\vec{\theta}=(t_\phi,t_g,t_{\rm{GVD}})^T$ determine the phase velocity, group velocity and group-velocity dispersion of a Fourier-limited pulse $E(t)=E_0 u(t)$.
We follow \cite{Jian:12}, and describe the temporal mode $u(t)$ in the frequency domain
\begin{equation}
    u(\omega)=\int u(t)e^{i\omega t}dt=\frac{1}{(2\pi\Delta^2\omega)^{1/4}}e^{-\frac{(\omega-\omega_0)^2}{4\Delta^2\omega}},
\end{equation}
where $\omega_0$ is the mean frequency and $\Delta^2 \omega$ its variance. In a dispersive medium, the pulse at a distance $L$ from the origin obtains a frequency-dependent phase and is given by
\begin{equation}
    E(\omega)=E_i(\omega) e^{ik(\omega)L}.
\end{equation}
Expanding $k(\omega)=\frac{n_\phi(\omega)\omega}{c}$ around the mean frequency $\omega_0$, we find
\begin{equation}
    E(\omega;\vec{\theta})=E_i(\omega)\exp[i(\omega_0t_\phi+(\omega-\omega_0)t_g+\frac{(\omega-\omega_0)^2}{\omega_0}t_{\rm{GVD}})]=E_0 u(\omega;\vec{\theta})
\end{equation}
where the parameters $\vec{\theta}=(t_\phi,t_g,t_{\rm{GVD}})^T$ are given by
\begin{align}
    t_{\phi} &= n_{\phi} (\omega_0) \frac{L}{c} , \notag \\
    t_{g} &= n_{g} (\omega_0) \frac{L}{c} = \left( n_{\phi} (\omega_0) + \omega_0 n'_{\phi} (\omega_0) \right) \frac{L}{c} , \notag \\
    t_{\text{GVD}} &= \omega_0 \left( n'_{\phi} (\omega_0) + \frac{\omega_0}{2} n''_{\phi} (\omega_0) \right) \frac{L}{c} .
\end{align}
The prime denotes the derivative to $\omega$. These parameters characterize the phase velocity, group velocity and broadening of the envelope of the pulse. Applying \autoref{eq:singlemodeQFIM}, we find for the QFIM
\begin{equation}
    F_Q(\hat{\rho})=4\begin{pmatrix}
        \omega_0^2F_Q^I&0&\Delta^2\omega F_Q^I\\
        0 & \Delta^2\omega N & 0\\
        \Delta^2\omega F_Q^I & 0 & \frac{(\Delta^2\omega)^2}{\omega_0^2}(F_Q^I+2N)
    \end{pmatrix},
\end{equation}
where $F_Q^I$ is given by \autoref{eq:QFIMfirstterm}. We see here the correlation between $t_\phi$ and $t_{\rm{GVD}}$, while $t_g$ is independent from the others. Here again, this can be understood from the detection mode calculated in \cite{Jian:12}. It is also shown that the parameters can be measured independently with a balanced homodyne detection scheme, if the mode of the local oscillator is carefully chosen. 
This consequently leads to a lowered precision of each parameter with respect to doing single parameter estimation. 

\section{Conclusions}
The fundamental limits of the estimation of multiple parameters encoded in the mode of light depend on how the mode changes with the parameter. In this work, the general expression for the QFIM for any mode parameter is found for any type of quantum state. Specifically, if the probe light is either single mode or has a strong mean field, every parameter corresponds to a specific mode, namely the derivative mode of the single populated mode or of the mean-field mode, respectively. The precision in single parameter estimation is determined by the the magnitude of the (non-normalized) derivative mode ($\propto(f^\alpha|f^\alpha)$) while in multiparameter estimation the correlation between parameters is determined by the overlap between the corresponding derivative modes ($\propto(f^\alpha|f^\beta)$).
While the quantum Cramér-Rao bound gives insights into the ultimate achievable precision, in multiparameter estimation it is not always saturable since the optimal measurements of different parameters can be fundamentally incompatible. We find that for mode parameters, this incompatibility is determined by the imaginary part of the overlap of the derivative modes, which can be directly linked to the non-commutivity of the quadratures of the derivative modes.
The results presented here can be used to calculate the precision on any mode parameter, the correlations between parameters and the compatibility of optimal measurements. Therefore the formalism  gives insights on how to optimize the precision of the estimation of correlated or uncorrelated parameters by carefully shaping the modes of the probe light and populating them with non-classical light. 

\section*{Acknowledgements}
This project has received funding from the European Defence Fund (EDF) under grant agreement 101103417 EDF-2021-DIS-RDIS-ADEQUADE. Funded by the European Union. Views and opinions expressed are however those of the author(s) only and do not necessarily reflect those of the European Union or the European Commission. Neither the European Union nor the granting authority can be held responsible for them.
This work was supported by the Fraunhofer Internal Programs under Grant No. Attract 40-09467. This work was supported by the project PID2023-152724NA-I00, with funding from MCIU/AEI/10.13039/501100011033 and FSE+ and by the project CNS2024-154818, with funding from MICIU/AEI/10.13039/501100011033. This work was funded by MCIN/AEI/10.13039/501100011033 and the European Union ‘NextGenerationEU’ PRTR fund [RYC2021-031094-I], by the Ministry of Economic Affairs and Digital Transformation of the Spanish Government through the QUANTUM ENIA Project call—QUANTUM SPAIN Project, by the European Union through the Recovery, Transformation and Resilience Plan—NextGenerationEU within the framework of the Digital Spain 2026 Agenda, and by the CSIC Interdisciplinary Thematic Platform (PTI+) on Quantum Technologies (PTI-QTEP+). This work was supported through the project CEX2023-001292-S funded by MCIU/AEI.

\bibliography{bibliography}

\end{document}